\documentclass[nobibnotes,twocolumn,10pt]{revtex4}
\usepackage[T1]{fontenc}
\usepackage{revsymb}
\usepackage{graphicx}
\usepackage{amsmath}
\usepackage{psfrag}
\newcommand{\bfi}{\begin{figure}[hbtp]}

\newcommand{\efi}{\end{figure}}

\newcommand{\dr}{\partial}
\newcommand{\beq}{\begin{equation}}
\newcommand{\eeq}{\end{equation}}
\newcommand{\beqar}{\begin{eqnarray}}
\newcommand{\eeqar}{\end{eqnarray}}

\begin{document}

\title{Asymptotic properties of wall-induced chaotic mixing in point vortex pairs.}
\author{Jean-R\'egis Angilella}

\affiliation
{Nancy-Universit\'e, LAEGO,
 rue du Doyen Roubault, 54501 Vand\oe uvre-les-Nancy, France}

\vskip.5cm
\begin{abstract}
{ The purpose of this work is to analyze the flow due to a potential  point vortex pair in the vicinity of a symmetry line (or "wall"), in order to understand why the presence of the wall, even far from the vortices, accelerates fluid mixing around the vortex pair. An asymptotic analysis, in the limit of large distances to the wall, allows to approximate the wall effect as a constant translation of the vortex pair parallel to the wall, plus a straining flow which induces a natural blinking vortex mechanism with period half the rotation period.
A Melnikov analysis of lagrangian particles, in the frame translating and rotating with the vortices, shows that   a homoclinic bifurcation  indeed occurs, so that the various   separatrices located near the vortex pair (and rotating with it) do not survive when a wall is present.  The thickness of the resulting inner stochastic layer  is estimated by using the separatrix map method, and is shown to scale like the  inverse of the squared distance to the wall. This estimation provides a lower-bound to the numerical thickness measured from either Poincar\'e sections or simulations of lagrangian particles transported by the exact potential velocity field in the laboratory frame. In addition, it is shown that the  outer homoclinic cycle, separating the vortices from the external (open) flow, is also perturbed from inside by the rotation of the vortex pair. As a consequence, a stochastic layer is shown to exist also in the vicinity of this cycle, allowing fluid exchange between the vortices and the outer flow. However, the thickness of this outer stochastic zone is observed to be much smaller than the one of the inner stochastic zone near vortices, as soon as the distance to the wall is large enough.  

}
\end{abstract}
\vskip1cm

\maketitle


\section{Introduction}

Vortical flows have remarkable transport properties which have been investigated in various contexts  in the past  decades. Indeed, vortices are   often referred to as "violent" or "singular" structures which induce large local velocity gradients which in turn significantly influence the transport of scalar or vector fields. For example,  steady two-dimensional isolated fixed vortices, whether pointwise or with a viscous core, have been shown to significantly influence the transport of passive scalars (Flohr \& Vassilicos \cite{Flohr1997}) or magnetic fields (Bajer \cite{Bajer1998}). If, in addition, the vortex oscillates, then the
 velocity field is unsteady, and this unsteadiness is likely to influence scalar mixing
 through some chaotic advection phenomenon (Aref \cite{Aref1984}
\cite{Aref2002}, Beigie {\em et al.} \cite{Beigie1994}). One of
the simplest unsteady vortex of this kind is the periodically translating isolated vortex, which can be thought of as a variation of the blinking vortex, and which has been shown to
induce fast mixing in chaotic regions (Wohnas \& Vassilicos \cite{Wonhas2001}).
In the case of vortex pairs the flow is also unsteady, since both vortices rotate around each other (provided the total strength is non-zero),
 but this unsteadiness is not
sufficient to induce chaotic mixing. Indeed, in the frame rotating with the vortex pair the flow is steady, and the motion of fluid points is an autonomous dynamical system with only
1 degree-of-freedom: their motion cannot be chaotic.

If, in addition, the two vortices move in the vicinity of a flat wall, then elementary vortex dynamics shows that both the distance between the vortices and their rotation velocity are no longer constant. Fluid point trajectories are therefore much more complex. It has been shown that chaotic advection indeed occurs in the case of leapfrogging vortex pairs (Pentek {\em et al.} \cite{Pentek1995}), which can be thought of as a vortex pair near a wall simulated by using the mirror method.  In the present paper we will investigate this phenomenon   analytically to show that chaos persists when the vortices are far from the wall, and to quantify this effect by calculating the thickness of the resulting stochastic layers.
  
We will use an asymptotic method to show that the basic ingredient  of chaos in the work by  Rom-Kedar {\em et al.} \cite{RomKedar1990}, namely the artificial
unsteady  straining flow applied to a counter-rotating vortex pair, is also at work in the case of a co-rotating vortex pair near a wall.
These authors show that counter-rotating  vortex pairs can  induce chaotic advection,
 provided they are superposed to an oscillating straining flow.
They observe that in the absence of the oscillating straining flow "the pair carries a constant body of fluid", whereas the oscillating straining flow forces the fluid to be "entrained and detrained from the neighbourhood of the vortices". 

The goal of the present paper is to show that,
for co-rotating vortex pairs, a wall-induced straining flow can make the two vortices oscillate to-and-fro with a period half the natural rotation period of the pair, and that this 
variation of the blinking vortex phenomenon triggers  chaotic mixing.
This result will be derived by using an asymptotic approach, in the
 limit $d_0 \ll L_0$, where $d_0$ is the order-of-magnitude of the distance 
between vortices, and $L_0$ is the distance to the wall.
This "inner" asymptotic velocity will approximate the flow in the vicinity of vortices, i.e. at distance $O(d_0)$ from them. Wall-induced chaotic mixing will then be analyzed by considering the breaking of the corresponding "inner" homoclinic cycles.

In addition, it will be shown that vortex rotation also perturbs, from inside, 
the outer homoclinic cycle
separating the vortex pair from the external open flow. This effect will be analysed by means of an
 "outer" asymptotic velocity field (in the same
 limit $d_0 \ll L_0$), designed to approximate the flow at distance $L_0$ from vortices.

The inner asymptotic model is depicted in section \ref{secasym}. Chaotic advection and mixing in this flow are investigated and discussed in section \ref{chaotadv}. We then focus on fluid exchange with the outer open flow by using the outer approximation (section \ref{ExtSep}). Because $d_0 \ll L_0$, 
the motion of vortices will be assumed to be regular throughout the paper (see Acheson \cite{Acheson2000}).

 \begin{figure}[htbp]
        \centering
                \includegraphics[width=0.4\textwidth]{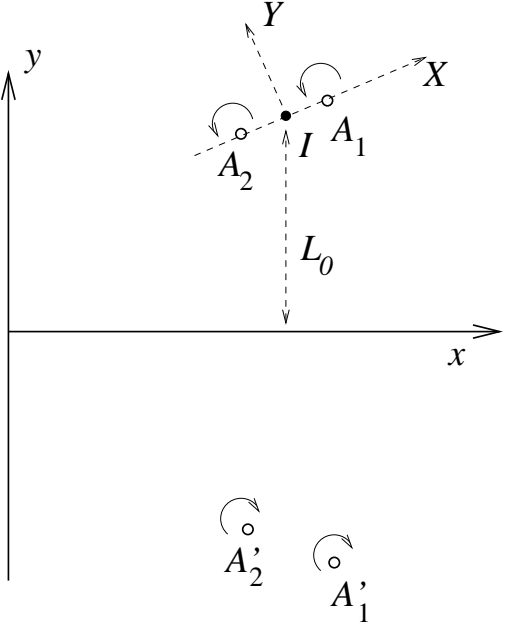}
        \caption{  Sketch of the vortex pair $(A_1,A_2)$, together with the two mirror vortices
 $(A'_1,A'_2)$.  }
        \label{coords}
\end{figure}

\section{Inner asymptotic velocity field near vortices}
\label{secasym}

We consider a pair of point vortices $(A_1,A_2)$  with equal strength $\Gamma$ moving in the plane $(O,x,y)$.
In the absence of any solid boundary  
both vortices  rotate around their centre point $I$ with an angular velocity
$
\Omega_0 = {\Gamma}/{4 \pi d_0^2}  ,
$
where $2d_0 = |A_1 A_2|$ is the (constant) distance between the vortices in this case.
In the presence of a  wall, located at $y=0$, the streamfunction is (inviscid fluid):
$$
\psi(x,y,t) = \sum_{i=1}^2 - \frac{\Gamma}{4 \pi} \mbox{Log} \left[ (x-x_i)^2 + (y-y_i)^2 \right]
$$
\beq
+ \sum_{i=1}^2 \frac{\Gamma}{4 \pi} \mbox{Log} \left[ (x-x_i)^2 + (y+y_i)^2 \right],
\label{psiexact}
\eeq
where  $(x_1,y_1)$ are the coordinates of point $A_1$, and $(x_2,y_2)$ are the coordinates of point $A_2$. The latter sum is the wall effect, and will be next simplified in the limit where the wall is far from the vortices  (see for example Appendix A of Rom-Kedar {\em et al.} \cite{RomKedar1990}).
Elementary vortex dynamics shows that:
\beq
\dot x_I = \frac{\Gamma L_0}{4 \pi}\left(\frac{1}{y_1 y_2}+\frac{4}{|z_1-\bar z_2|^2} \right),
\label{xIexact}
\eeq
$$
y_I = constant = L_0,
$$
in all cases, with $z_k=x_k+i y_k$. By setting $y_k=y_I+\bar y_k$, with $\bar y_k = O(d_0) \ll L_0$, and expanding  (\ref{xIexact}) we get:
$$
\dot x_I = \dot x^0_I \left( 1+O(\varepsilon^2) \right)  ,
$$
with
$
 \dot x^0_I = {\Gamma  }/{2 \pi L_0},
$
and
$
\varepsilon = {d_0}/{L_0} \ll 1.
$
This shows that, to leading order, $I$ moves under the effect of a single mirror vortex with strength $-2 \Gamma$.
If non-dimensionalized by $\Omega_0$ and $d_0$, the $x$ velocity  of $I$ reads:
\beq
\frac{\dot x_I}{\Omega_0 d_0} = 2\varepsilon + O(\varepsilon^3),
\label{xIapprox}
\eeq
so that the vortex pair moves at constant speed up to order 3.
In order to exploit further the fact that  $d_0 \ll L_0$, we consider positions ($x,y$) located in the vicinity of the vortex pair, and normalize the variables as:
$
x^* = (x-x^0_I)/d_0, \quad y^* = (y-y_I)/d_0, $
and $x_i^* = (x_i-x^0_I)/d_0, \quad y_i^* = (y_i-y_I)/d_0,
$
where $x^0_I = \Gamma t / 2\pi L_0$.
Also, we set $t^* = \Omega_0 t$ and define the non-dimensional streamfunction in the frame translating at the velocity of $I$:
\beq
\psi^*(x^*,y^*,t^*) = \frac{\psi - \dot x^0_I \, (y-y_I)}{\Omega_0 d_0^2}  .
\label{psistarcomplet}
\eeq
Then, by expanding the streamfunction in the limit $\varepsilon \to 0$,  we get
 (removing the stars):
$$
\psi(x,y,t) = \sum_{i=1}^2 - \mbox{Log}\left[ (x-x_i)^2 + (y-y_i)^2 \right]
$$
\beq
 + \frac{\varepsilon^2}{2}  (x^2 - y^2 )  + O(\varepsilon^3).
\label{psiapprox}
\eeq
Terms of order $O(\varepsilon)$ have
cancelled-out with $\dot x^0_I$, as expected. Note that this flow does not correspond to a uniform flow $- \dot x^0_I$ at infinity (in contrast with Eq. (\ref{psistarcomplet})), as it is only an approximation of the streamfunction in the {\it vicinity} of the vortices.  

\subsection{Approximate vortex motion}

\begin{figure}[htbp]
        \centering
                \includegraphics[width=0.5\textwidth]{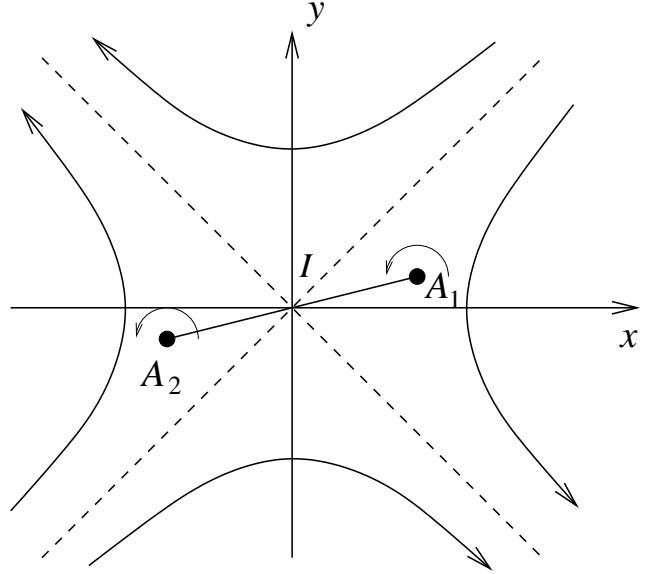}
        \caption{Sketch of the stretching flow induced by the wall in the reference frame translating with $I$. This flow stretches and compresses the segment $A_1 A_2$ twice during each complete rotation of the vortex pair.}
        \label{StretchCompressAB}
\end{figure}

We therefore observe that the effect of the wall can be approximated by a stretching flow with axes $\vec e_x \pm \vec e_y$ (where $\vec e_x$ and $\vec e_y$ are the unit vectors attached to axes $x$ and $y$ respectively). 
Note that this flow is very close to the one investigated by  Carton {\em et al.} \cite{Carton2002} and  Maze {\em et al.} \cite{Maze2004}.
The line $A_1 A_2$ joining the two vortices is stretched and compressed periodically during the motion. It will be shown later that the period of this stretching/compression is half the period of the vortices without wall.
The dynamics of vortex $A_1=(x_1,y_1)$ in this simplified flow is:
\beqar
\label{dotx1}
  \dot x_1 &=&   -  2 \frac{y_1-y_2}{ 4 r^2 } -   \frac{\varepsilon^2}{2}  (   y_1 - y_2) , \\
  \dot y_1  &=&     2 \frac{x_1-x_2}{ 4 r^2 } -   \frac{\varepsilon^2}{2}     ( x_1-x_2)  ,
\label{dotx2}
\eeqar
with $2r(t) = |A_1 A_2|$. By writing the motion equation of $A_2=(x_2,y_2)$ we are led to:
$
(\dot x_1 + \dot x_2)/2 = 0 = (\dot y_1 + \dot y_2)/2.
$
This shows that, to this order, the centre point $I$ is indeed fixed, in agreement with equation (\ref{xIapprox}).  The latter equation was also expected: indeed, as noticed in the introduction,  $y_I$ is known to remain constant whatever the distance to the wall.
The motion of the vortices can then be investigated further by setting
$
x_1(t) = r(t) \cos \theta(t),\quad y_1(t) = r(t) \sin \theta(t),
$
so that equations (\ref{dotx1}) and (\ref{dotx2}), with $x_2 = -x_1$ and $y_2=-y_1$ lead  to:
\beqar
\dot r &=& -  \varepsilon^2 \, r \, \sin 2 \theta,\\
\dot \theta &=& \frac{1}{r^2} - \varepsilon^2 \cos 2 \theta.
\eeqar
Looking for an asymptotic solution in powers of $\varepsilon$, we get (setting $r(0)=1+\varepsilon^2/2$ and $\theta(0)=0$):
\beqar
r(t) &=& 1 +\frac{\varepsilon^2}{2} \cos 2 t,\\
\theta(t) &=& t - \varepsilon^2 \sin 2 t,
\eeqar  
and the vortex coordinates in the reference frame translating with $I$ then read:
\beqar
\label{x1asym}
x_1(t) &=& \cos t + \varepsilon^2 \left( \frac{\cos t}{2}  \cos 2 t + \sin t \sin 2 t \right)\\
y_1(t) &=& \sin t + \varepsilon^2 \left( \frac{\sin t}{2}  \cos 2 t - \cos t \sin 2 t \right).
\label{y1asym}
\eeqar
We therefore observe that the wall, when far from the vortex pair, creates an oscillation of the distance between the two vortices with period $\pi$ (half the period of the isolated vortex pair). Also, the rotation velocity of the vortices
around $I$ is affected by a perturbation  with period $\pi$. This periodic forcing corresponds to the effect of the straining flow $\varepsilon^2 (-y,-x)$ appearing in the velocity field, which stretches and compresses the vortex pair twice during each rotation (see Fig. \ref{StretchCompressAB}).

 In the following lines we write $\vec x=(x,y)$, and the vortex position is also written in vector form:
$\vec x_i(t)  = \vec x_i^{(0)} + \varepsilon^2 \vec x_i^{(2)}$, with the components of the various vectors given by (\ref{x1asym})-(\ref{y1asym}). Then the streamfunction in the frame translating with $I$, written in powers of $\varepsilon$, reads:
\beq
\psi_a(\vec x,t) = \sum_{i=1}^2 \psi_0\left(\vec x - \vec x_i(t)\right) + \varepsilon^2 \psi_2(\vec x),
\eeq
with $\psi_0(\vec x)  = -\mbox{Log}(x^2+y^2)$ and $\psi_2(\vec x) = (x^2-y^2)/2$. It can be expanded as:
$$
\psi_a(\vec x,t) = \sum_{i=1}^2 \left[\psi_0\left(\vec x - \vec x_i^{(0)}(t)\right)
- \varepsilon^2 \vec x_i^{(2)} . \nabla \psi_0|_{\vec x - \vec x_i^{(0)}} \right]
$$
\beq
\label{psia}
 + \varepsilon^2 \psi_2(\vec x).
\eeq
This streamfunction corresponds to the flow induced by  the two vortices and the wall, in the reference frame translating with $I$, and is valid only in the vicinity of the vortex pair $A_1 A_2$.

\subsection{Flow in the rotating frame}

It is very convenient to investigate the motion of lagrangian particles in the reference frame translating with $I$ {\it and rotating} at speed $\Omega_0$, that is the angular 
velocity of the vortex pair in the absence of a wall. Let $\vec e_X,\vec e_Y$ denote an orthogonal basis attached to this frame, with $X,Y$ the corresponding coordinates. Assuming $\vec e_X=\vec e_x$ and $\vec e_Y=\vec e_y$
at $t=0$ we have:
$
x = X \cos t - Y \sin t
$
and $y = X \sin t + Y \cos t$. Then, in this reference frame, and with these coordinates, the streamfunction reads
$$
\psi_r(X,Y,t) = \psi_a\left(x(X,Y,t),y(X,Y,t),t\right) + \frac{1}{2}  (X^2+Y^2),
$$
where the last term corresponds to the entrainment velocity of the rotating frame. By using (\ref{psia}) we obtain, after some algebra:
\beq
\psi_r(X,Y,t) = \psi_{r0}(X,Y)+\varepsilon^2\psi_{r2}(X,Y,t),
\label{psir}
\eeq
where
\beq
\psi_{r0}(X,Y,t) = -\mbox{Log}|Z^2-1|^2 + \frac{1}{2} |Z|^2,
\eeq
and
$$
\psi_{r2}(X,Y,t) =- \frac{2}{|Z^2-1|^2} [ (Y^2-X^2+1) \cos 2 t
$$
\beq
+ 4 X Y \sin 2 t ]+ \frac{1}{2} \left[ (X^2-Y^2) \cos 2 t - 2 X Y \sin 2 t \right],
\label{psir1}
\eeq
where $Z=X+iY$. The leading order streamfunction corresponds  to the steady flow in the rotating frame
in the absence of wall, the $\varepsilon^2$ terms are the wall effect.

\section{Wall-induced chaotic advection near vortices}
\label{chaotadv}

\subsection{Splitting of the inner homoclinic cycle}

\begin{figure}[htbp]
        \centering
                \includegraphics[width=0.5\textwidth]{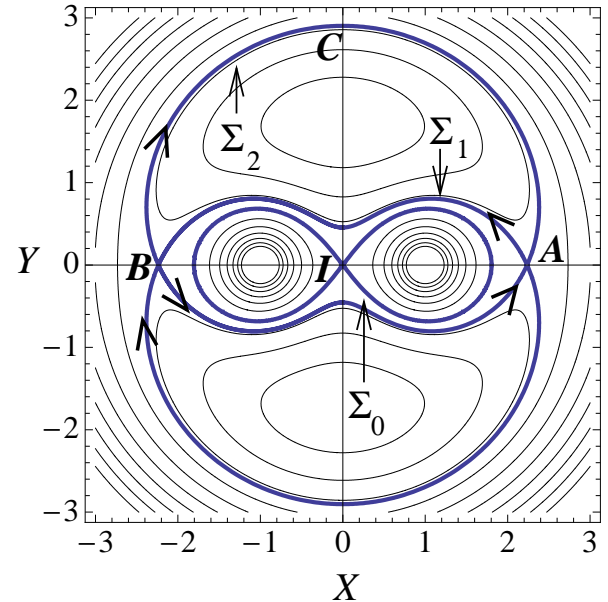}
        \caption{ Streamlines of the leading-order flow in the rotating frame, in the vicinity of
 the vortices (inner asymptotic field).}
        \label{LdCAvec2SeparatricesFIG}
\end{figure}

\begin{figure}[htbp]
        \centering
                \includegraphics[width=0.5\textwidth]{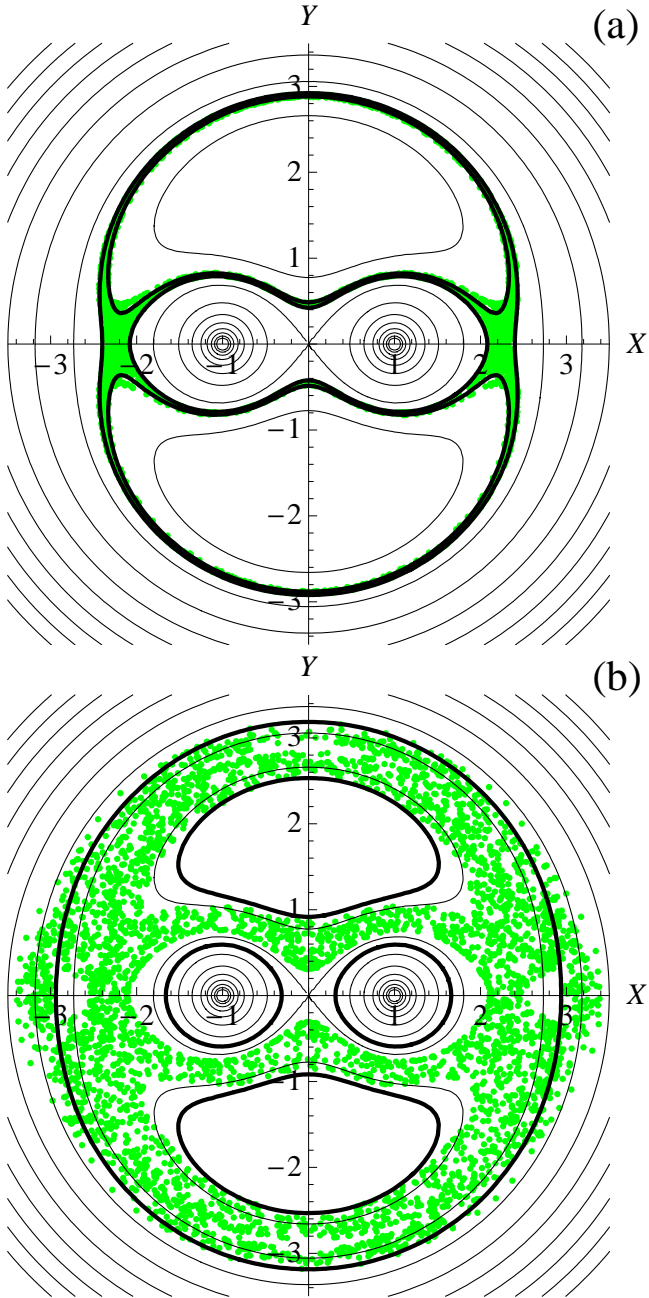}
        \caption{ Poincar\'e sections of 20 particles, together with the theoretical boundary of
 the inner stochastic zone (black thick line). Graph (a): $\varepsilon = 0.05$, graph (b): $\varepsilon = 0.2$.}
        \label{secpoincEpsi0.05et0.2}
\end{figure}

In order to investigate the occurrence of chaotic advection in the flow (\ref{psir})
we consider fluid points, the dynamics of which reads:
\beq
\frac{d\vec X}{dt} =   \frac{\dr \psi_r}{\dr Y} \vec e_X - \frac{\dr \psi_r}{\dr X} \vec e_Y,
\label{dynsyst}
\eeq
where $\vec X(t) = (X,Y)$ is the position of the fluid point.
This is a perturbed hamiltonian system,
 the phase portrait of which is, to leading order, 
the same as the one of Fig.\ 2a of Angilella \cite{Angilella2010}, which is reproduced here (Fig.\ \ref{LdCAvec2SeparatricesFIG}) for
clarity. This widely used phase portrait has two homoclinic cycles ($\Sigma_1 \cup  \Sigma_2$ and its
symmetric in the lower half-plane), and two homoclinic trajectories ($\Sigma_0$ and its
symmetric in the left half-plane).   Let $\vec q_i(t)$ be a solution of the leading-order fluid points dynamics:
$$
\frac{d\vec q_i}{dt} =  \frac{\dr \psi_{r0}}{\dr Y} \vec e_X - \frac{\dr \psi_{r0}}{\dr X} \vec e_Y,
$$
with initial condition on a separatrix $\Sigma_i$ ($i=0,1,2$). Then the occurrence of chaos in the fluid point dynamics under the effect of the $\varepsilon^2$ terms is related to the occurrence of simple zeros in the Melnikov function of the separatrix $\Sigma_i$:
 \beq
M_i(t_0) = \int_{-\infty}^{\infty}   \left[\psi_{r0}(\vec q_i(t)) , \psi_{r2}(\vec q_i(t),t+t_0) \right] dt,
\label{Mit0}
\eeq
where $t_0$ is the starting time of the Poincar\'e section of the perturbed tracer dynamics, and $[.,.]$ is the Poisson bracket.
By choosing $\vec q_i(0)$   on the "middle-point" of $\Sigma_i$ (i.e. the intersection between the separatrix and $OY$ for $\Sigma_1$ and $\Sigma_2$, and $OX$, $X\not = 0$, for $\Sigma_0$), then the coordinates of $\vec q_i(t)$ are either  odd or even functions, and by using the various symmetries of the velocity field, one can get rid of the integral of odd functions of $t$ in (\ref{Mit0}). We are led to
$
M_i(t_0) = \alpha_i \sin(2 t_0),
$
where the amplitudes of the Melnikov functions $\alpha_i$ are purely numerical constants.
They are calculated  by solving the $\vec q_i(t)$'s
 numerically on the three separatrices (in the upper half-plane for $\Sigma_{1,2}$ and in the right half-plane for $\Sigma_{0}$): $\alpha_0 \simeq -0.58$, $\alpha_1 \simeq -0.89$, $\alpha_2 \simeq 7.31$. We
therefore conclude that the two homoclinic cycles and the two homoclinic trajectories of
Fig.\ \ref{LdCAvec2SeparatricesFIG} do not survive when a wall is present. This is confirmed by the Poincar\'e sections shown in Fig.\ \ref{secpoincEpsi0.05et0.2}. A stochastic zone  indeed exists in the vicinity of the separatrices, and grows with $\varepsilon$. 
It will be called "inner" stochastic zone in the following, as it exists in the very vicinity of the vortices (i.e. at distance $O(d_0)$ from them).
The typical scale of this zone, and its boundaries, are investigated in the next section.  

\subsection{Characterization of the inner stochastic zone}

\begin{figure}[htbp]
        \centering
                \includegraphics[width=0.5\textwidth]{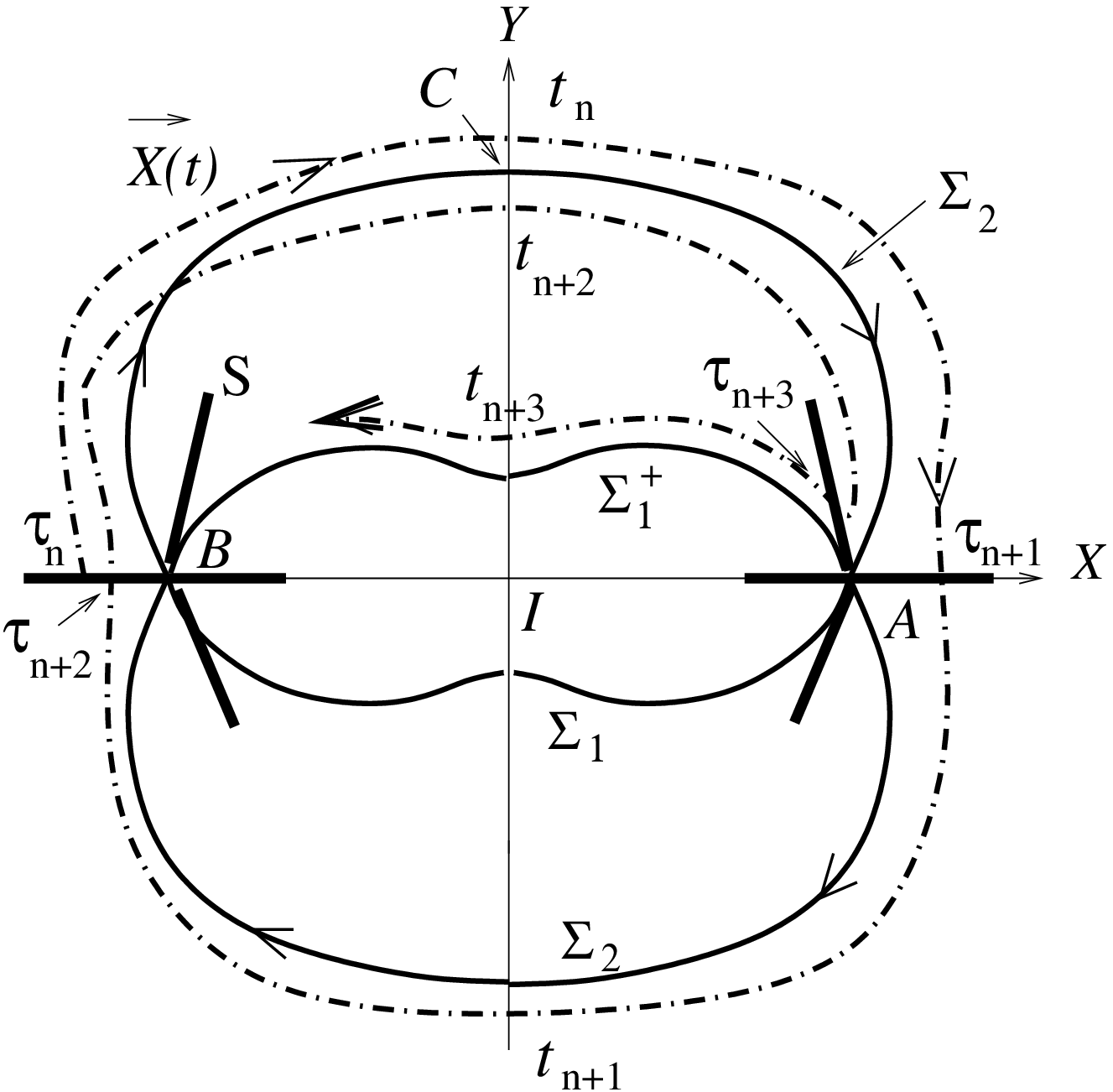}
        \caption{Sketch of the separatrix map used to calculate the  boundary of
the inner stochastic layer.}
        \label{MethodeSepMap}
\end{figure}

In order to quantify mixing in the vicinity of the vortices we derive a semi-analytic
expression for the equation of the border of the stochastic layer.
The thickness of this zone grows with $\varepsilon$, and can be calculated from the above Melnikov functions (Chirikov \cite{Chirikov1979}, Weiss \& Knobloch \cite{Weiss1989},  Rom-Kedar \cite{RomKedar1994}, Kuznetsov \& Zaslavsky \cite{Kuznetsov1998}, Trueba \& Baltanas \cite{Trueba2003}, Balasuriya \cite{Balasuriya2005}).
We build the separatrix map of $\Sigma_i$ for $i \in \{1,2\}$ (Fig.\ \ref{MethodeSepMap})
  by considering
 a fluid point
$\vec X(t)$  running under the effect of the perturbed flow (\ref{psir}). This point crosses   the   axis $IY$ at discrete times $t_n$.   In addition, following Weiss \& Knobloch \cite{Weiss1989}, we define the set $S$ composed of the hyperbolic points  $A$ and $B$, and of   segments forming two symmetric stars near $A$ and $B$ (Fig.\ \ref{MethodeSepMap}). These stars have
 two branches parallel to $IX$,  and two other branches lying between $\Sigma_1$ and $\Sigma_2$. The detailed shape of the $S$ is of no importance to construct the separatrix map, provided
the trajectory of $\vec X(t)$ remains close to the separatrices. In addition we call $\tau_n$ the time at which the trajectory of $\vec X(t)$ intersects $S$ just prior time $t_n$, that is: $\tau_n < t_n < \tau_{n+1}$.

 Let $H_n$ denote the value of the unperturbed streamfunction $\psi_{r0}$ at $\vec X(\tau_n )$. The variation of "energy" $H_n$
between two consecutive crossings of $S$ is:
$$
 \Delta H_n=  H_{n+1}-H_n = \varepsilon^2 \int_{\tau_{n}}^{\tau_{n+1}} \left[\psi_{r0}(\vec X(t)) , \psi_{r2}(\vec X(t),t) \right] dt.
$$
If the trajectory is close enough to a separatrix (Chirikov \cite{Chirikov1979}) , say $\Sigma_i$, then we have $\vec X(t) \simeq \vec q_{i}(t-t_n)$, where $\vec q_{i}$, parametrizing $\Sigma_i$, has been introduced in the previous section, and is such that  $\vec q_{i\pm}(0)$ belongs to the vertical axis $IY$.
 Also, by noticing   that $\tau_n \ll t_n \ll \tau_{n+1} $, we obtain:
\beq
 \Delta H_n \simeq  \varepsilon^2   M_i (t_n),
\label{sepmapH}
\eeq
where $M_i$ are the Melnikov functions of separatrices $\Sigma_i$ calculated in the above section. The time lag $t_{n+1}-t_n$ can be approximated by the half-period of the unperturbed orbit\cite{Chirikov1979} $\psi_{r0}=H_{n+1}$, so that we set:
\beq
t_{n+1}=t_n + T(H_{n+1})/2.
\label{sepmaptau}
\eeq
Equations (\ref{sepmapH}) and (\ref{sepmaptau}) approximate the separatrix map in the vicinity of the homoclinic cycle $\Sigma_1 \cup \Sigma_2$.
To our knowledge, no exact analytical expression for the period $T(H)$ is known. However, following Kuznetsov \& Zaslavsky \cite{Kuznetsov1998}, one can notice that the dynamics of fluid points is very slow in the vicinity of point $A$, so that an approximate expression of $T(H)$ can be found by expanding the streamfunction in the vicinity of $A$. We then obtain hyperbolic streamlines for this simplified dynamics, and $T(H)/2$ is approximated as the time required for a fluid point running on an hyperbolic streamline to pass over $A$. We are led to (see Appendix \ref{AppA}):
\beq
T(H)   \simeq -\beta \mbox{Log} |H_A-H|,
\label{Tapprox}
\eeq
with $\beta = 4/\sqrt 5$.


The amplitude of the rate of change of the time interval $d(t_{n+1}-t_n)/dt_n$  can then be calculated from equations (\ref{sepmapH}) and (\ref{sepmapH}):
\beq
\mbox{Max}\left|\frac{d t_{n+1}}{d t_n} - 1\right| 
= \varepsilon^2\, {\beta  |\alpha_i |}/{|H_A-H_{n+1}|}.
\eeq
 By writing that this quantity is equal to unity on the border of the stochastic
layer  (Rom-Kedar \cite{RomKedar1994}) we obtain the  energy  $H_s$ corresponding to this boundary:
\beq
|H_{s}-H_A| =  \beta  |\alpha_i | \, \varepsilon^2.
\label{Hs}
\eeq
The boundary of  the stochastic layer around $\Sigma_2$ can then be obtained by setting $i=2$ in the above formula.  
The curve $\psi_{r0}=H_{s}$ has been plotted on the numerical Poincar\'e sections of Fig.\ \ref{secpoincEpsi0.05et0.2} for comparison (thick black line), and we observe
that, in spite of the rough approximations of the model, this curve corresponds to the  border of the stochastic zone with an acceptable accuracy.

\begin{figure}[htbp]
        \centering
                \includegraphics[width=0.5\textwidth]{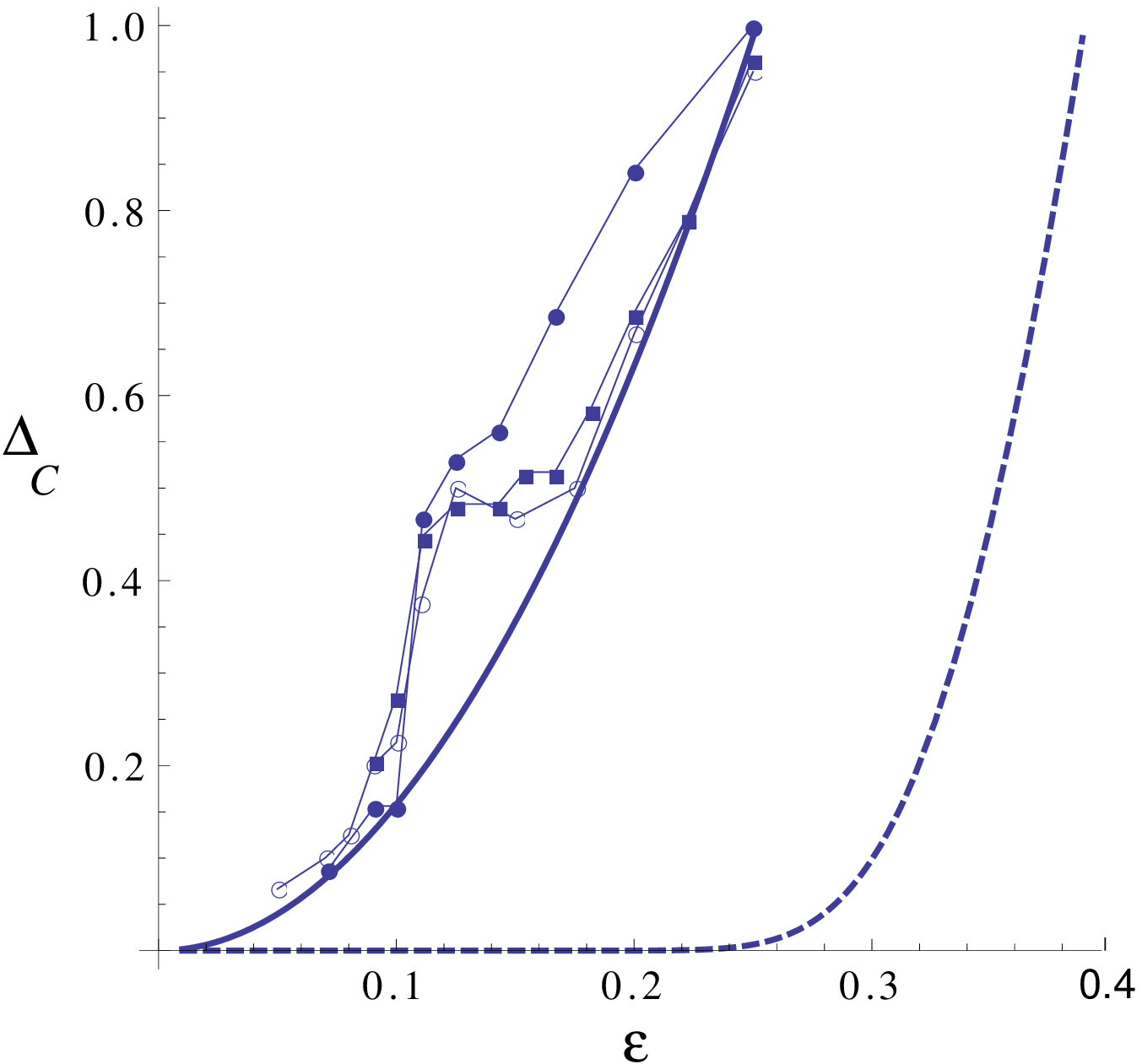}
        \caption{Thickness of the inner stochastic zone at point $C$ of $\Sigma_2$, versus
 $\varepsilon=d_0/L_0$, obtained from the separatrix map (equation (\ref{rextC}), thick
 solid line) and estimated from numerical computations: Poincar\'e sections (empty
 circles), particle cloud using the asymptotic velocity in the rotating frame (squares),
 particle cloud using the exact potential velocity in the laboratory frame (filled circles). 
The dashed line is the theoretical thickness of the outer stochastic zone near the 
separatrix $S_1 S_2$ discussed in section \ref{ExtSep}.}
        \label{L0ThickFIG}
\end{figure}

Result (\ref{Hs}) can also be used to determine the distance $r_{C}$ between the
 border of the stochastic layer and point $C$, which can be thought of as the
 half-thickness of the stochastic layer at $C$. Indeed, by writing that $\psi_{r0}(\vec X_C + r_{C} \vec e_Y) = H_{s}$, and using a first-order Taylor expansion (assuming that $r_{C}$ is small):
\beq
r_{C} =    \frac{\beta  \alpha_2}{\gamma} \, \varepsilon^2,
\label{rextC}
\eeq
where $\gamma = \dr \psi_{r0}/\dr Y(C) \simeq 1.67$. The thickness of the stochastic layer
of $\Sigma_2$ at $C$ is therefore:
$
\Delta_C = 2 r_{C} \simeq 15.8 \epsilon^2.
$
This formula is compared to measurements of the thickness $\Delta_C$  obtained from numerical computations (Fig.\ \ref{L0ThickFIG}). Three kinds of computations
are used here: Poincar\'e sections, particle clouds computed by using the asymptotic velocity in the rotating frame, and particle clouds computed by using the
 exact potential velocity (\ref{psiexact})
in the laboratory frame. Point $C$ has been chosen for these
measurements because  the thickness is  well-defined there. We observe that the order-of-magnitude of $\Delta_C$ is well predicted by the theory, especially if $\varepsilon$ is small. For larger $\varepsilon$ discrepancies of about 100\% appear, and these errors are known to be due to resonances occurring in the fluid point dynamics (Luo \& Han \cite{Luo2001}, Trueba \& Baltanas \cite{Trueba2003}) and to regular islands embedded in the stochastic layer. In spite of these discrepancies we can conclude that the theoretical prediction obtained from the separatrix map  gives an acceptable lower bound to the thickness of the stochastic zone.

The typical size of the stochastic layer can also be estimated in the vicinity of the hyperbolic points $A$ (and $B$) , by writing that $\psi_{r0}(\vec X_A + r_{A} \vec e_X) = H_{s}$, where $r_{A}$ is the distance from $A$ to the external border of the stochastic zone, towards  $\vec e_X$. To obtain $r_{A}$ we then use a second-order Taylor expansion (assuming that $r_{A}$ is small). Indeed, because $A$ is a stagnation point of the unperturbed flow, the gradient of $\psi_{r0}$ vanished there, and we are left with:
\beq
r_{A} =   2 \left(\frac{\beta  \alpha_2}{5}\right)^{1/2} \, \varepsilon.
\label{rextA}
\eeq
The typical size of the stochastic layer is therefore larger in the vicinity of $A$ (and $B$) than anywhere else, in the limit where $\varepsilon$ is small. This is a rather general result which is a consequence of the fact that $A$ and $B$ are stagnation points \cite{Kuznetsov1998}. Here also we expect (\ref{rextA})  to provide only a lower bound of the actual $r_{A}$, since resonances might occur and make the thickness even bigger.

\subsection{Comparison with transport by the exact potential velocity field}
    
The asymptotic model presented in the above sections shows that some chaotic mixing must occur in the vicinity of a vortex pair as soon as a wall is present in the surroundings of the two vortices. In order to check the reliability of these calculations, which are valid only if the wall is  far from the vortices, we have performed some calculations of fluid points transport  by using the exact potential velocity field (\ref{psiexact}) (set dimensionless by using $\Omega_0$ and $d_0$),  in the laboratory frame.

\begin{figure}[htbp]
        \centering
                \includegraphics[width=0.4\textwidth]{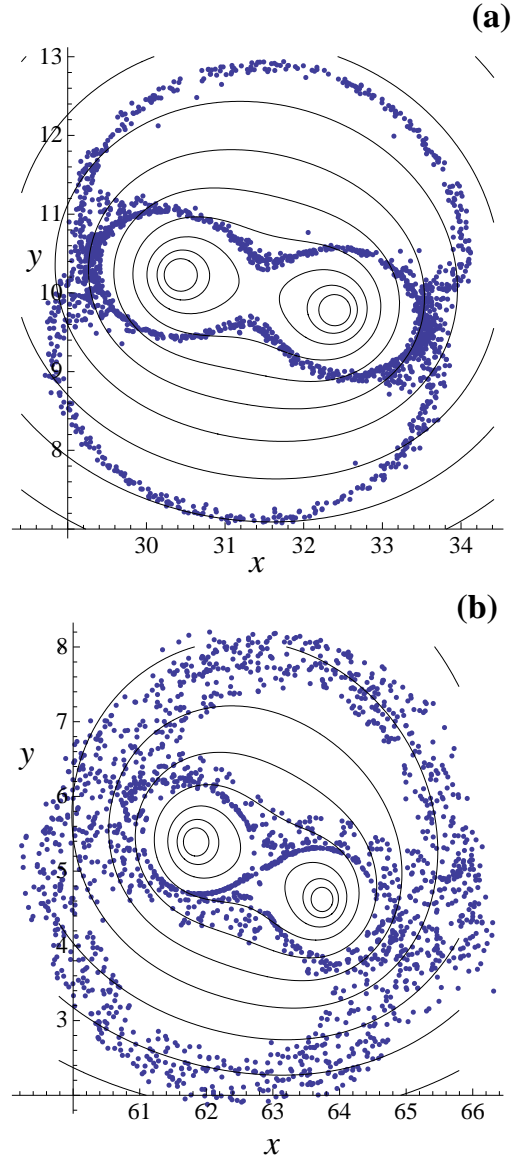}
        \caption{Particle clouds at $t=50 \pi$, computed in the laboratory frame by using
 the exact potential velocity field, for $L_0 = 10 d_0$ ($\varepsilon = 0.1$) (a), 
and $L_0 = 5 d_0$ ($\varepsilon = 0.2$).  Solid lines are streamlines.}
        \label{DNStraceurs_teq50piL0eq5FIG}
\end{figure}

Fig.\ \ref{DNStraceurs_teq50piL0eq5FIG} shows two particle clouds corresponding to $L_0 = 10 d_0$ and $5 d_0$ (that is $\varepsilon = 0.1$ and $0.2$, respectively). Because our goal is to visualize fluid point mixing in the stochastic zone, the initial cloud is taken to be a small spot of size $0.1\times 0.1$ centered at $I$ (which is initially located at $(0,0)$).
The initial positions of the vortices are $(x_1,y_1)=(1+\varepsilon^2/2,1/\varepsilon)$ and
$(x_2,y_2)=(-1-\varepsilon^2/2,1/\varepsilon)$.   The final time of the three simulations is $t=50 \pi$, and the final location of the centre point is very close to the asymptotic value  $(x_I,y_I) \approx (2 \varepsilon \, t,L_0)$ in both cases. We observe that some particles collect on a structure which is very close to the separatrices $\Sigma_i$ described above, as a consequence of the existence of a stochastic zone in the vicinity of these separatrices. The thickness of the stochastic zone is of the order of the theoretical one, as shown in Fig.\ \ref{L0ThickFIG} (filled circles).

\section{Mixing with the outer open flow}
\label{ExtSep}

\begin{figure}[htbp]
        \centering
   \includegraphics[width=0.5\textwidth]{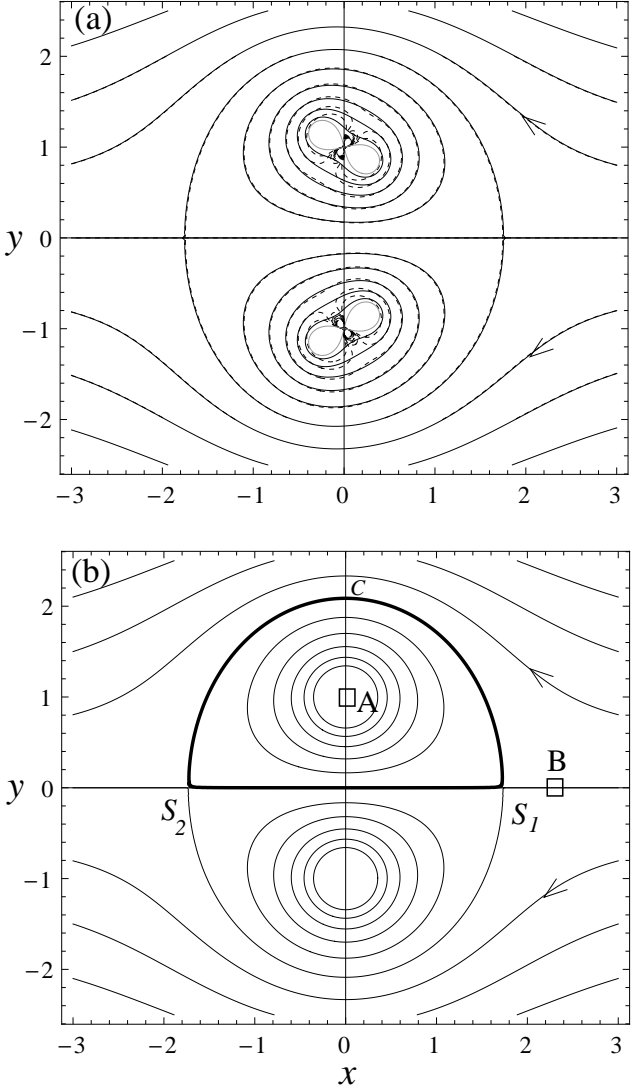}
        \caption{Graph (a): streamlines in the frame translating with the vortices,
 when $\varepsilon=0.3$. Both the real ($y>0$) and mirror ($y<0$) vortices are shown ; the wall
 corresponds to $y=0$.
Solid lines: exact four-vortex potential flow (\ref{psiexact}). Dashed lines: outer
asymptotic streamfunction (\ref{psiext}). Graph (b): leading-order flow ($\varepsilon = 0$) in the same frame. $S_1$ and $S_2$ are saddle points located at $(\pm \sqrt 3,0)$. The thick line is the homoclinic cycle $S_1 S_2$.  (Lengths have been rescaled by $L_0$, stars have been removed.)}
        \label{PsiRefTranslAsymS1S2Epsi0.3}
\end{figure}

The results of the above sections show that chaotic advection can occur 
in the vicinity of the vortices, due to the presence of the wall which induces a periodic unsteadiness in the vortex motion. 
However, the inner asymptotic model (valid only at distance $O(d_0)$ from vortices) 
does not bring any information about fluid mixing with the
"external flow" which carries fluid elements from the far field. 
Fig.\ \ref{PsiRefTranslAsymS1S2Epsi0.3}(a) shows the complete flow (solid lines), obtained from the exact potential solution (\ref{psiexact}), in the frame translating with   vortices. 
Both the real and mirror vortices are shown.
Lengths and velocities have been normalized by $L_0$ and $\Gamma/L_0$ respectively, which are relevant scales for the external flow, and which will be used to derive an outer asymptotic velocity field below.  The vortex system is separated from the rest of the flow by a homoclinic cycle  which can be thought of as the boundary between closed and open streamlines. The breaking of this homoclinic cycle is obviously an important mechanism for fluid mixing between the vortex system and the fluid outside
(Pentek {\em et al.} \cite{Pentek1995}). This homoclinic cycle will be called the
 "outer cycle"  in the following, as opposed to the so-called "inner cycle" $\Sigma_2 \cup 
\Sigma_1$.

Because the flow is two-dimensional, the only candidate for the splitting of this outer cycle is the rotation of the vortices around each other, which induces an unsteady perturbation to the homoclinic cycle $S_1 S_2$. This bifurcation is investigated in the next paragraphs.

\subsection{Asymptotic analysis of the outer homoclinic cycle}
\label{splitext}

As noticed above, $L_0$ and $\Gamma$ being the relevant scales of the external flow, we non-dimensionalize the streamfunction (\ref{psiexact}) by using $L_0$ for lengths and $\Gamma/4\pi L_0$ for velocities.  Also, we consider the reference frame translating at the velocity $\dot x_I^0 = {\Gamma  }/{2 \pi L_0}$ of a vortex with strength $2 \Gamma$, placed at distance $L_0$ from the wall. The non-dimensional streamfunction of the flow induced by the four vortices with strengths $\Gamma,\Gamma,-\Gamma,-\Gamma$, in this frame, is:
\beq
\psi_*(\vec X_*) = \frac{\psi - \dot x_I^0 \, y}{\Gamma/4\pi},
\eeq 
where $\psi(x,y,t)$ is the streamfunction of the flow observed in the laboratory frame, and stars indicate non-dimensional variables.
Still assuming that $\varepsilon = d_0 / L_0 \ll 1$, and using Eqs.\ (\ref{x1asym}) and (\ref{y1asym}) for the vortex motion, the streamfunction can be expanded as:
$$
\psi_*(\vec X_*) \simeq \psi_{0*}(\vec X_*) + \varepsilon^2 \psi_{2c}(\vec X_*) \cos \frac{2 t_*} {\varepsilon^2}
$$
\beq
+ \varepsilon^2 \psi_{2s}(\vec X_*) \sin \frac{2 t_*}{\varepsilon^2}   + O(\varepsilon^4),
\label{psiext}
\eeq
where $ \psi_{0*}(x_*,y_*)=2\log[(x_*^2+y_*^2+1+2y_*)/(x_*^2+y_*^2+1-2y_*)]-2y_*$ is the streamfunction of a simple dipole centered at (0,0) (i.e. single vortex plus its mirror vortex, see Fig.\  \ref{PsiRefTranslAsymS1S2Epsi0.3}(b)), and the $\varepsilon^2$ terms manifest the fact that we have a vortex {\it pair} with strength $\Gamma+\Gamma$, instead of a single vortex with strength $2 \Gamma$, so that the flow is unsteady. The streamfunctions $\psi_{2c}$ and $\psi_{2s}$
are steady, rational functions of $x_*$ and $y_*$ only.

With the time scale used here, the rotation of the vortex pair is fast, so that the flow (\ref{psiext}) is a {\it rapidly perturbed} dipolar flow. Fig.\ \ref{PsiRefTranslAsymS1S2Epsi0.3}(a) shows this streamfunction when $\varepsilon = 0.3$ (dashed line), together with the exact four-vortex streamfunction already discussed 
(solid line). The two flows are very close, except in the vicinity of both pairs, as expected.  The $O(\varepsilon^3)$ terms in (\ref{psiext}) are identically zero (for symmetry reasons), and this is why the agreement is good even if $\varepsilon$ is not very small here.  
For $\varepsilon \le 0.2$ dashed lines and solid lines (not shown here) are indistinguishable, except in the very vicinity of the vortices. 

Particle motion can therefore be approximated by a perturbed hamiltonian system, with hamiltonian $ \psi_{0*}$.
The Melnikov function attached to the cycle $S_1 S_2$ reads (removing the stars):
$$
M_{ext}(t_0) = \int_{-\infty}^{\infty}   \left[\psi_{0}(\vec q(t)) , \psi_{2c}(\vec q(t))\right] \cos (2\frac{t+t_0}{\varepsilon^2})  dt 
$$
\beq
+ \int_{-\infty}^{\infty}   \left[\psi_{0}(\vec q(t)), \psi_{2s}(\vec q(t))\right] \sin (2\frac{t+t_0}{\varepsilon^2})  dt,
\label{Mt0ext}
\eeq
where $\vec q(t)$ runs on the unperturbed cycle, and is chosen such that $\vec q(0)$ coincides with point C of Fig.\ \ref{PsiRefTranslAsymS1S2Epsi0.3}(b). This Melnikov function
approximates (to order $\varepsilon^2$) the "energy" difference  $\psi_0(\vec X(\tau_2)) - \psi_0(\vec X(\tau_1))$, where $\vec X(t)$ is the trajectory of a fluid point advected by the perturbed flow, with a trajectory very close the separatrix $S_1 S_2$. The times $\tau_1$ and $\tau_2$ correspond to the particle passing near the saddle points $S_1$ and $S_2$.
Because our perturbation is rapid, the amplitude of the Melnikov function will now depend on $\varepsilon$ (Gelfreich \cite{Gelfreich1997}). The integrands in the above integrals are unknown analytically,
they are however localized around zero and decay exponentially. We have chosen to approximate them with a mixed polynomial/gaussian expression of the form 
 $P(t)/\cosh^2(\delta \, t)$, where $P$ is a fitted polynomial and $\delta$ is a fitted 
 constant. 
The integrals can then be obtained analytically, and we are led to $M_{ext}(t_0) =  A(\varepsilon) \sin ({2 t_0}/ {\varepsilon^2}) $, with:
\beq
A(\varepsilon) =  \frac{\gamma}{\varepsilon^2} {e^{-\pi / \delta \varepsilon^2}},
\eeq
where $\delta$ and $\gamma$ are purely numerical constants depending on the fitting procedure: $\gamma \approx 18$, $\delta \approx 5$.
Note that this formula depends on the form chosen to approximate our integrands. This is therefore a numerical result, rather than a truely analytical one, being understood that other fittings might not significantly affect the numerical values of the amplitude of the Melnikov function.
Because $A(\varepsilon) > 0$ for all $\varepsilon$, and $M_{ext}(t_0)$ always has simple zeros when $t_0$ varies, portions of fluid will be exchanged between the interior of the cycle $S_1 S_2$, and the fluid outside. However, the amplitude is so small as soon as $\varepsilon \le 0.2$ (i.e. $A(0.2) < 7\, 10^{-5}$), that mixing through the cycle $S_1 S_2$ should be hardly visible for such small $\varepsilon$'s.  
In contrast, when $\varepsilon$ increases above 0.2, $A(\varepsilon)$ quickly reaches values of order unity. By constructing the separatrix map of $S_1 S_2$, one can derive 
the theoretical thickness $\Delta$ of the stochastic layer surrounding the outer homoclinic cycle.
In the vicinity of point C of Fig.\ \ref{PsiRefTranslAsymS1S2Epsi0.3}(b),
we are led to $\Delta_C \approx 0.5 A(\varepsilon)$. This quantity has been plotted in Fig.\  \ref{L0ThickFIG} (dashed line). We indeed observe that the thickness is exponentially small when $\varepsilon \le 0.2$ ($L_0 \ge 5 d_0$), and quickly reaches values of order unity as soon as $\varepsilon$ is larger than about 0.25. [Note that, even though the diagram
suggests a sharp increase of $\Delta_C$, there is no critical $\varepsilon$ here: the Melnikov function has simple zeros for all  $\varepsilon$.]
 Fig.\ \ref{L0ThickFIG} suggests that the vortex pair
might be {\em apparently} separated from the outer open flow when $L_0  \ge 5 d_0$, since the whole vortex system will be surrounded by an extremely thin stochastic zone. Also, the thickness of this stochastic zone should quickly reach a value of order unity as soon as $L_0 \le 3 d_0$.
 These two points are checked numerically in the next paragraph.

\subsection{Numerical verification}

 \begin{figure}[htbp]
        \centering
                \includegraphics[width=0.33\textwidth]{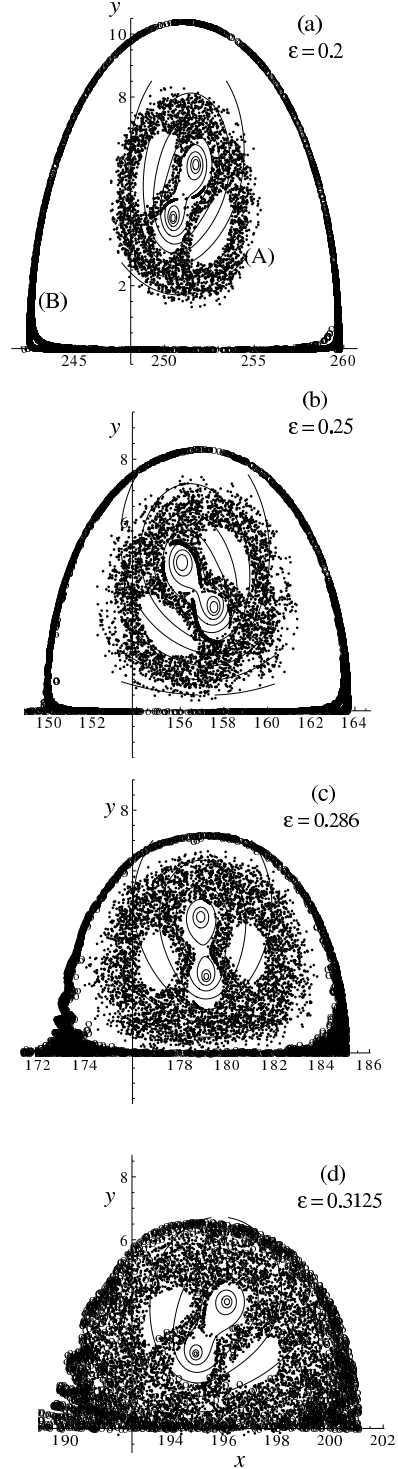}
        \caption{The role of the inner and outer separatrices in chaotic mixing. 
The four graphs show particle clouds in the laboratory frame at $t=100 \pi$, obtained by 
using the exact four-vortex potential solution. 
 Particles marked with black dots have been injected near the upper vortices at $t=0$ 
(in box A of Fig.\ \ref{PsiRefTranslAsymS1S2Epsi0.3}(b)). Those marked with empty circles have been injected outside the cycle $S_1S_2$, ahead of vortices
 (in box B of Fig.\ \ref{PsiRefTranslAsymS1S2Epsi0.3}(b)).
Graph (a): $\varepsilon \simeq 0.2$ ($L_0/d_0 = 5$). Graph (b): $\varepsilon \simeq 0.25$ ($L_0/d_0 = 4$).
Graph (c): $\varepsilon \simeq 0.286$ ($L_0/d_0 = 3.5$). Graph (d): $\varepsilon \simeq 0.3125$ ($L_0/d_0 = 3.2$).    Solid lines are streamlines.}
        \label{nuage5L0eq4et3.2tf314}
\end{figure}

Computations of particle clouds in the laboratory frame, using the exact four-vortex 
potential solution (\ref{psiexact}),  have been performed. Particles have been injected at $t=0$ near the vortices (in box A of Fig.\ \ref{PsiRefTranslAsymS1S2Epsi0.3}(b)). Also, to check whether portions of fluid could enter within the cycle $S_1 S_2$, we have injected another set of particles outside the cycle $S_1S_2$ (in box B of Fig.\ \ref{PsiRefTranslAsymS1S2Epsi0.3}(b)). These simulations confirm the significant
 difference between the thickness of the inner cycle and the one of the outer cycle,
as suggested by the theoretical curves of Fig.\ \ref{L0ThickFIG}.

Indeed, Fig.\ \ref{nuage5L0eq4et3.2tf314}(a) shows the two particle clouds at $t=100 \pi$, when $L_0/d_0 = 5$ ($\varepsilon = 0.2$). 
Units are the same as in the simulations of section \ref{chaotadv}, that is: $d_0$ for lengths and $\Omega_0$ for times. 
Particles marked with black dots are those which have been injected in box A (they are also indicated with a (A)), and those marked with empty circles have been injected in box B (they are also indicated with a (B)).   
Even though these computations have been conducted in the laboratory frame, the homoclinic cycles appearing in the rotating frame (section \ref{chaotadv}) and in the translating frame (section \ref{splitext}), naturaly appear here. Particles injected near vortices wander in the vicinity of the inner cycle $\Sigma_2 \cup \Sigma_1$ investigated in section \ref{chaotadv},  whereas those injected ahead of the vortices move in the vicinity of the outer 
cycle $S_1 S_2$. Some particles penetrate within this cycle, in agreement with the fact that  the Melnikov function has simple zeros. 
When $L_0/d_0 = 4$ (Fig.\ \ref{nuage5L0eq4et3.2tf314}(b)) the thickness of the inner stochastic zone around the cycle $\Sigma_2$ is larger, whereas the 
   thickness of the outer 
stochastic zone around the cycle $S_1 S_2$ does not appear to be 
very different from case (a), in agreement with the theoretical thickness calculated in the previous section (dashed line of Fig.\ \ref{L0ThickFIG}).
When $L_0/d_0 = 3.5$ (Fig.\ \ref{nuage5L0eq4et3.2tf314}(c)) the thickness of the outer stochastic zone  is slightly larger than in case (b), but the two clouds are still separated. 
For  $L_0/d_0 = 3.2$  ($\varepsilon = 0.3125$, Fig.\ \ref{nuage5L0eq4et3.2tf314}(d)) the two particle clouds mix significantly, in agreement with the fact that the amplitude of the Melnikov function is of order unity there. However, the thickness of the outer 
stochastic zone cannot be measured from these graphs and compared to the theoretical value of $\Delta_C$: this is due to the fact that particles injected in box B now mix with those injected in box A. These simulations confirm, however, that the outer stochastic zone remains negligible compared to the inner one as long as 
$L_0/d_0$ is larger than about 4, so that the vortex pair might exchange very little fluid
with the outer  flow in this case.

\section{Conclusion  }

We have analyzed how point vortex pairs, which are closed flow structures in the absence of walls,
exchange portions of fluid with their surroundings as soon as a wall, even far from the vortices, is present.
This is not a new result, since Pentek {\em et al.} \cite{Pentek1995} had shown that chaotic advection is present in leapfrogging vortex pairs, but the asymptotic approach used here
shows that   the mechanism at work is a combination of the 
stretching flow induced by the wall and of the unsteadiness due to vortex rotation. Our work has therefore some common features with the one of Rom-Kedar {\em et al.} \cite{RomKedar1990}, even though the flow is different. Relevant quantities,  like the thickness of   stochastic layers, have been calculated by using Melnikov functions, as done in previous works for other flows (e.g.
Kuznetsov \& Zaslavsky \cite{Kuznetsov1998}, Trueba \& Baltanas \cite{Trueba2003}).  

 The breaking of the inner separatrices, located in the very vicinity of the vortex pair and
 rotating with them,
induces a stochastic layer  the thickness of which scales like the squared inverse distance to the wall.
 The rotation of vortices around each other can also affect
 the outer 
homoclinic cycle $S_1 S_2$  separating the vortex system and the rest of the flow domain. 
  The thickness of this outer stochastic layer has been shown to be
negligible as long as $\varepsilon \le 0.25$, so that the vortex pairs
might be {\em apparently} separated from the open flow zone when $L_0 \ge 4 d_0$. 
However, this thickness rapidly jumps to unity for larger  $\varepsilon$'s. As a consequence, significant portions of fluid coming from infinity can
penetrate within this cycle and mix with the fluid located near vortices.  
 
One could argue that the values of $\varepsilon$ considered in the
 various computations presented here 
are not very small, so that the comparison with the asymptotic results might 
be done with care. However, the asymptotic analyses show that the
 leading-order wall effect scales like $\varepsilon^2$. The approximation is especially 
accurate in the case of the outer separatrix (section \ref{ExtSep}) since the velocity
truncation error scales like  $\varepsilon^4$ there. 

The generalization to vortices with unequal strengths can be done by using the same asymptotic approach when the wall is far from the vortices, provided the sum of the two strengths is non-zero.
Here also we obtain that  the    center of vorticity  $I$  (strength-weighted averaged position) moves at constant speed up to order 3 and that the wall induces a stretching flow, in the frame translating with $I$, which triggers a blinking-vortex mechanism. One can check that the various separatrices  do not survive when a wall is present in this case also.

Finally, it should be noted that the "wall" analyzed in this work is in fact a symmetry line, rather than a wall, as the inviscid fluid considered  here slips on solid boundaries. The generalization of these results to real fluids, using the Navier-Stokes equations together with no-slip boundary conditions, must be done with care. Indeed, it is well known that vortices, even with a large Reynolds number, induce counter-rotating secondary vortices when a wall is present. This vorticity, which is initially located inside the wall boundary layer, can contaminate the flow and interact with the primary vortices, leading to a complex vortex system, very different from the ideal one investigated here.  
 
\vskip.1cm
\noindent
{\bf Acknowledgement}
\vskip.1cm
The author would like to thank A. Motter for fruitful discussions at  Northwestern University, and T. Nizkaya for her comments.
\appendix
\section{Calculation of the period near the separatrix}
\label{AppA}

\begin{figure}[htbp]
        \centering
                \includegraphics[width=0.4\textwidth]{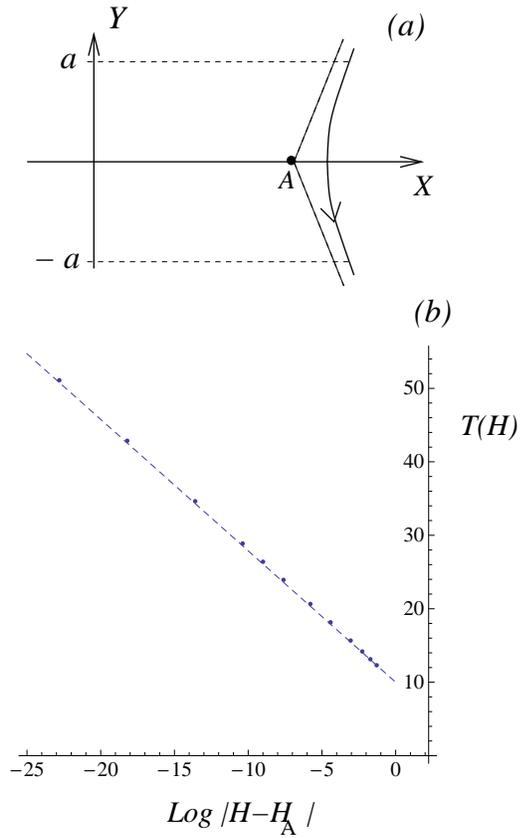}
\caption{Graph (a): linearized flow in the vicinity of  stagnation point A. Graph (b): comparison between the theoretical period of trajectories near the separatrix (Eq. (\ref{Tapprox}), dashed line) and the numerical one (dots). }
        \label{TFIG}
\end{figure}

In the vicinity of saddle point $A$ the unperturbed streamfunction can be expanded as:
$$
H = \psi_{r0}(\vec X) \simeq H_A + \frac{1}{2}  (\vec X-\vec X_A).\nabla^2 \psi_{r0} . (\vec X-\vec X_A) 
$$
\beq  
\simeq H_A -\frac{Y^2}{4} + \frac{5}{4} (X-X_A)^2.
\label{psiapproxapp}
\eeq
It corresponds to open hyperbolic trajectories like the one shown in Fig.\ \ref{TFIG}(a). For such a trajectory the time required to pass near point $A$ is very large if $\delta H=|H-H_A|$ is small, and this time is a good approximation of the half-period of trajectories around the homoclinic cycle \cite{Kuznetsov1998}. We therefore write
$$
T(H)/2 \simeq \int_{a}^{-a} \frac{d Y}{V_Y} 
$$
where $a$ is an arbitrary initial height of the fluid point, which has very little effect on $T(H)$ when the trajectory is close to the separatrix. The vertical velocity is $V_Y = -(5/2) (X-X_A)$, where $X$ is given by (\ref{psiapproxapp}), and we get:
$$
T(H) = -\frac{8}{\sqrt 5} \mbox{Log} \frac{2  { \delta H}^{1/2}}{a+(4 \delta H + a^2)^{1/2} }.
$$
In the limit $\delta H \to 0$, keeping $a$ fixed, we obtain Eq. (\ref{Tapprox}). 
Fig.\
\ref{TFIG}(b) shows a plot of Eq. (\ref{Tapprox}) together with a numerical measurement of $T(H)$, and we observe that the agreement is excellent.
 
\end{document}